\documentclass[fleqn,usenatbib]{mnras}

\usepackage{newtxtext,newtxmath}
\usepackage[T1]{fontenc}

\DeclareRobustCommand{\VAN}[3]{#2}
\let\VANthebibliography\thebibliography
\def\thebibliography{\DeclareRobustCommand{\VAN}[3]{##3}\VANthebibliography}

\usepackage{graphicx}	% Including figure files
\usepackage{amsmath}	% Advanced maths commands
\usepackage{tablefootnote}
\usepackage{longtable}
\usepackage[dvipsnames]{xcolor}
\usepackage{float}
\newcommand{\ngc}{N$_{\rm GC}$}
\newcommand{\mgc}{M$_{\rm GC}$}
\newcommand{\mt}{M$_{\rm T}$}

\title[Globular clusters in low mass galaxies]{Revisiting the relation between the number of globular clusters and galaxy mass for low mass galaxies}

\author[Dennis Zaritsky]{Dennis Zaritsky$^{1}$\thanks{E-mail: dennis.zaritsky@gmail.com}
\\\\
$^{1}$Steward Observatory and Department of Astronomy, University of Arizona, Tucson, AZ 85719, USA\\
}

\date{Accepted XXX. Received YYY; in original form ZZZ}

\pubyear{2022}

\begin{document}
\label{firstpage}
\pagerange{\pageref{firstpage}--\pageref{lastpage}}
\maketitle
%\author[0000-0002-5177-727X]{Dennis Zaritsky}

\begin{abstract}
Using a new method to estimate total galaxy mass (\mt) and two samples of low luminosity galaxies containing measurements of the number of globular clusters (GCs) per galaxy (\ngc), we revisit the \ngc-\mt\ relation using a total of 203 galaxies, 157 of which have \mt $\ \le 10^{10}$ M$_\odot$. We find that the relation is nearly linear, \ngc $\propto$ \mt$^{0.92\pm0.08}$ down to at least \mt $\ \sim 10^{8.75}$ M$_\odot$. Because the relationship extends to galaxies that average less than one GC per galaxy and to a mass range in which mergers are relatively rare, the relationship cannot be solely an emergent property of hierarchical galaxy formation. The character of the radial GC distribution in low mass galaxies, and the lack of mergers at these galaxy masses, also appears to challenge models in which the GCs form in central, dissipatively concentrated high-density, high-pressure regions and are then scattered to large radius. The slight difference between the fitted power-law exponent and a value of one, leaves room for a shallow \mt-dependent variation in the mean mass per GC that would allow the relation between total mass in GCs and \mt\ to be linear.
\end{abstract}


%%%%%%%%%%%%%%%%%%%%%%%%%%%%%%%%%%%%%%%%%%%%%%%%%%

%%%%%%%%%%%%%%%%% BODY OF PAPER %%%%%%%%%%%%%%%%%%

\section{Introduction}
\label{sec:intro}

One of the more intriguing relationships among galaxy properties is that between the number of globular clusters (\ngc), or total mass contained therein (M$_{\rm GC}$), and total galaxy mass, M$_{\rm T}$ \citep{blakeslee,georgiev,harris13,hudson,forbes16,harris17,forbes18,burkert}. Such a relationship suggests that most GCs form primarily in ways disconnected to how galaxies form the bulk of their stars \cite[cf.,][]{spitler,hudson}, which in turn implies distinct star formation modes during galaxy evolution.

It is tempting to speculate that the \ngc-\mt\ and \mgc-\mt\ relations provide a critical clue to GC formation. However, generic
models where GC formation occurs predominantly at high redshift, in progenitor dark matter halos that are aggregated by hierarchical assembly into the galaxies we see today, can begin to reproduce the observed trend without resorting to exotic theories \citep{mbk,el-badry,choksi,burkert,bastian20}. In fact, \cite{el-badry} demonstrated that even when GCs are assigned randomly to progenitor halos, the galaxies eventually formed from those halos satisfy the relations. The 'fog' of hierarchical assembly apparently obscures our view of GC formation.

However, those same studies demonstrated that this interpretation would be challenged if the relationships extend to low mass galaxies, as some had already suggested to be the case \citep{hudson,zar16,harris17}, where mergers are less common and galaxies do not build up primarily through multiple hierarchical events \citep{fitts,martin}. Indeed,  \cite{forbes18} extended the \mgc-\mt\ relation to masses well below where the fiducial model prediction of \cite{el-badry} deviates from linearity and \cite{bastian20} emphasised that they predict a turnover in the \mgc-\mt\ relation. This is an important tension that merits further investigation, with implications for models of galaxy evolution and GC formation. As illustrated by \cite{burkert} and \cite{bastian20}, there is a compelling case for precisely determining the nature of the relationship between GCs and galaxy mass as far down the galaxy mass function as possible. 

The principal obstacles we face in extending and refining the \ngc\ or \mgc\ vs. \mt\ relations in low mass galaxies are the small numbers of GCs per galaxy at these galaxy masses and the paucity of mass measurements for such galaxies. For example, the \cite{forbes18} sample includes only 22 galaxies with M$_{\rm T} < 10^{10}$ M$_\odot$, where 10$^{10}$ M$_\odot$ is roughly the lower mass limit of previous studies \citep[e.g.,][]{spitler} and is one example of a hypothesised threshold mass below which galaxies do not form GCs \citep{el-badry}.
Fortunately, two
recent studies \citep{forbes20,carlsten} provide measurements of \ngc\ for large samples of low luminosity galaxies.
Unfortunately, there are yet no corresponding \mt\ measurements for these samples and it is difficult to measure internal kinematics for such a large number of faint, low surface brightness galaxies. This paucity of data results in state-of-the-art statistical studies focusing on the relation between N$_{\rm GC}$ or M$_{\rm GC}$ and M$_*$ instead of M$_{\rm T}$ \citep[e.g.,][]{eadie}. 

Faced with this challenge, we propose to obtain \mt\ estimates for these galaxies
using a new method that relies on galaxy scaling relations and, after doing so, examine the low mass behaviour of the \ngc-\mt\ relation. Although our approach is manifestly less precise, and potentially less accurate, than evaluating \mt\ from spectroscopic kinematic measurements, the small numbers of GCs per galaxy ($\lesssim 1$) for the lowest mass galaxies in these samples inherently limits the precision with which any galaxy can ever be placed on the \ngc-\mt\ relation. Therefore, precise \mt\ measurements offer little return on investment given the high observational cost of measuring the internal kinematics of these galaxies. Rather than working with a few galaxies where one axis in this space is inherently imprecise, we propose working with many galaxies where both axes are imprecise and recovering precision through averaging. We focus on \ngc-\mt\ rather than \mgc-\mt\ to stay closer to the observations, particularly for these galaxies where the mean GC mass is not well established. In \S2 we briefly describe the published literature data we are using. In \S3 we present our approach to estimating \mt\ and supporting evidence for the accuracy of the approach on average. In \S4 we discuss the results and summarise in \S5. We adopt WMAP9 cosmological parameter \citep{wmap} and a solar V-band absolute magnitude of 4.81 \citep{willmer}.

\section{The Data}
\label{sec:data}

We use two sources of \ngc\ measurements for low luminosity galaxies \citep{forbes20,carlsten}. The \cite{forbes20} study examines the GC populations of low surface brightness galaxies in the Coma cluster, with a focus on ultra-diffuse galaxies (those with $r_e>$ 1.5 kpc). They compiled measurements of \ngc\ for 76 galaxies from other studies \citep{vdk17,lim, amorisco18} within the footprint of the Subaru imaging they use to measure galaxy properties
\citep{alabi} and complement their galaxy structure measurements with an additional 9 from \cite{lim}. Here we use only the 76 systems in common between \cite{forbes20} and \cite{alabi} to ensure homogeneity among the structural parameters we utilise in the scaling relations used to derive galaxy masses.

The details of the completeness corrections used to correct both for those GCs that are below the detection limit and for those that lie beyond the adopted search radius are critical when comparing \ngc\ measurements among studies. At the distance of the Coma cluster, the corrections can often be greater than a factor of two \citep[e.g.,][]{vdk16} and highly uncertain. \cite{teymoor2} use deep {\sl HST} imaging to derive more precise constraints on the GC luminosity function and radial distribution for 6 Coma cluster ultra-diffuse galaxies. They find a luminosity function that is consistent with what was previously assumed and in agreement with that for GCs in other dwarf galaxies, but a radial distribution that is consistent with that of the stars and is thus different than that typically assumed, resulting in significantly revised, smaller completeness corrections. 
To reconcile the \ngc\ values compiled by \cite{forbes20} to these newer findings, we use the 5 galaxies in common between the \cite{forbes20} and  \cite{teymoor2} samples, to derive a median multiplicative correction of 0.27 for the \cite{forbes20} \ngc\ values. However, we caution that because the \cite{forbes20} sample is itself a compilation of \ngc\ values from different studies, a single recalibration may be insufficient. Unfortunately, we do not have enough overlap with each of the different studies incorporated to correct each independently.

If nothing else, this attempted recalibration illustrates the degree to which completeness corrections can affect the overall normalisation of \ngc. It is the largest source of uncertainty in the comparison among studies or to theoretical models. For further reference, had we chosen to correct by the mean rather than the median ratio then the 'corrected' \ngc\ values would be nearly two-thirds larger than what we are actually adopting. An error in this correction will affect an entire sample similarly and, therefore, will not affect conclusions regarding the linearity of the \ngc-\mt\ relation within the sample. However, it will result in offsets among studies, which is why having results from at least two different ones provides a valuable check. 

The second study we use is that of \cite{carlsten}, which presents \ngc, complemented by \cite{carlsten21} for the other necessary parameters for 145 low luminosity early-type galaxies in the Local Volume. The imaging data are sufficiently deep that they reach below the peak of the GC luminosity function. The GC radial distribution is extensively examined and discussed. For both of these reasons, and the fact that we have no overlapping measurements, we do not apply any multiplicative correction to their \ngc\ measurements. Because of the larger size and the homogeneity of the \ngc\ measurements, this is our preferred sample, but as we will show our results from the two samples are consistent. 

In both samples there are some galaxies with N$_{\rm GC} \le 0$ due to statistical background subtraction. The uncertainties at the smallest values of N$_{\rm GC}$ are dominated by the large background correction rather than the Poisson statistics of N$_{\rm GC}$ and therefore still closer to Gaussian in nature. These uncertainties are also significantly larger than those introduced by the completeness corrections (magnitude limit and radial distribution) if the GC populations of the low mass systems are not dramatically different than those of brighter systems where these corrections have been calibrated. As such, we do not anticipate a measurement bias near N$_{\rm GC}=0$ toward positive values of N$_{\rm GC}$.

\section{Estimating Total Mass}
\label{sec:fm}
In a series of papers \citep{FM,FM-LG,FM-structure,FM-clusters}, we presented an extension of the Fundamental Plane formalism \citep{djorgovski,dressler} that applies to the entire family of galaxies. Particularly relevant to the current discussion, the revised scaling relation applies to low mass stellar systems \citep{FM-LG,FM-clusters} and to ultra-diffuse galaxies \citep{zar17}. In the last of those studies, we exploited the relationship between the half-light radius, $r_e$, the surface brightness within $r_e$, $I_e$, and a measure of the internal kinematics, $V$, to estimate $V$, from which we then recovered the mass-to-light ratio within $r_e$, $\Upsilon_e$, and hence the mass within that same radius. Specifically, there are two equations that are empirically calibrated:
\begin{equation}
\log r_e = 2\log {\rm V} - \log I_e - \log \Upsilon_e - 0.8,
\label{eq:fm}
\end{equation}
\begin{equation}
\begin{split}
\log \Upsilon_e  &=   0.24\ ({\log {\rm V}})^2 + 0.12\ (\log I_e)^2  - 0.32\ {\log {\rm V}} \\
& \hspace{50pt} - 0.83\ {\log I_e} - 0.02\ \log {\rm V}I_e + A,
\end{split}
\label{eq:m2l}
\end{equation}
where the coefficients in Eq. \ref{eq:m2l} were obtained by fitting to a large sample of galaxies \citep{zaritsky12}. We specifically call out the constant $A$ and discuss it further below.
The kinematic term, 
V, is defined to be the combination of the line-of-sight velocity dispersion, $\sigma_v$, and the inclination corrected rotation speed, v$_{r}$, V$\equiv \sqrt{\sigma_v^2 + {\rm v}_{r}^2/2}$. These two equations can be rewritten as one and numerically solved for $V$ when one is given $r_e$ and $I_e$. In this specific version of the equations, $r_e$ is in units of kpc, $I_e$ in M$_\odot$ pc$^{-2}$, V in km sec$^{-1}$, and $\Upsilon_e$ in solar units.

\begin{figure}
\includegraphics[scale=0.5]{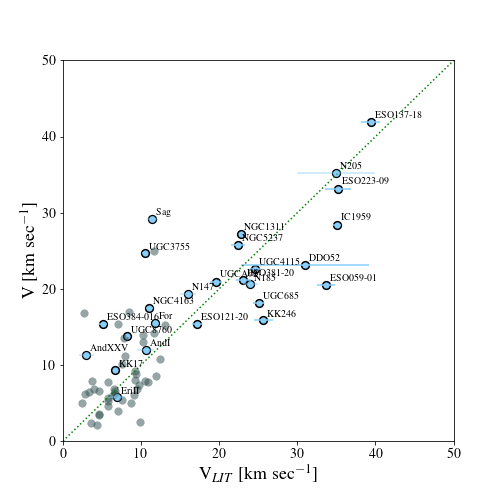}
\caption{Comparison of $V$ values derived using our formalism for two samples and those measured in the literature. Those galaxies with labels are from the \protect\cite{forbes18} sample, 
while the unlabelled ones are a set of Local Group dwarf galaxies compiled by M. Collins from a set of references \protect\citep{tollerud12,tollerud13,mcc,collins13,collins14,martin16}, excluding the three that \protect\cite{collins14} concluded 
are tidally distorted. The line is the 1:1 relation.}
\label{fig:sigma_comp}
\end{figure}

The constant $A$ was also previously determined using measured values of $r_e$, $V$, and $I_e$ for the same large  sample of galaxies. However, we adjust $A$ slightly here to provide better agreement between our calculated values of \mt\ and those available from \cite{forbes18}. We set $A = 1.57$, as opposed to the previous value of 1.49, and present the comparison between our estimated values of $V$ and the spectroscopically measured ones for the \cite{forbes18} sample and for other available low luminosity stellar systems in Figure \ref{fig:sigma_comp}. Although the scatter is significant, and the predicted $V$ for any one system can be catastrophically incorrect (wrong by more than a factor of 2), on average the estimates are accurate and can be used to recover the mean properties of galaxies. We will discuss further below the impact of using the previously determined value of $A = 1.49$ on our final results, but it principally affects our normalisation of the \ngc-\mt\ relation.

The use of the scaling relation only provides an estimate of the mass interior to $r_e$. To calculate the halo mass, \mt, we subtract the contribution to the mass interior to $r_e$ from stars within $r_e$ and determine the parameters of an NFW profile \citep{navarro} that reproduces the remaining mass within $r_e$. We do this by adopting the mean relation between concentration and mass from \cite{maccio}, using GalPy \citep[][http://github.com/jobovy/galpy]{bovy} to evaluate the enclosed mass at $r_e$, and iterating. For the best fit model, we evaluate
the halo virial mass, which we calculate for the adopted cosmology and a redshift of 0.01 to correspond to an overdensity of 346 relative to the matter density \citep[cf.][]{bryan}. The stellar contribution interior to $r_e$ is estimated using a V-band stellar mass-to-light ratio of 1.2 \citep{mcgaugh} for the \cite{forbes20} sample or the catalogued stellar mass for the \cite{carlsten} sample. To the calculated halo virial mass, we then add back the baryonic mass using the universal baryon fraction and recover \mt. We apply this methodology only to galaxies for which the stellar contribution to the mass within $r_e$ is fractionally small ($<$ 0.25 of the enclosed mass) to avoid highly uncertain values of the dark matter contribution within $r_e$ and mitigate the effects of ignoring possible adiabatic contraction. This criterion is responsible for the rejection of no galaxies from the \cite{forbes20} sample and 18 galaxies from the \cite{carlsten} sample. 

An interesting complication that we have sidestepped is that of scatter in the halo mass-concentration relation.
As \cite{maccio} show, there is significant scatter ($>0.1$ dex) measured in the relation as derived from simulations, which we have not accounted for in our mass estimation. Ignoring the scatter may, for a subtle reason, be the correct approach. Our estimates of the internal kinematics of these galaxies is based on  scaling relations, which also sidestep variations among individual galaxies and provide a `typical' velocity for each galaxy. As such, the velocities we try to fit to do not include the effect of variations in the concentration of the individual galaxies and we conclude that we should then estimate the masses using the mean concentration-halo mass relation.

\begin{figure}
\includegraphics[scale=0.5]{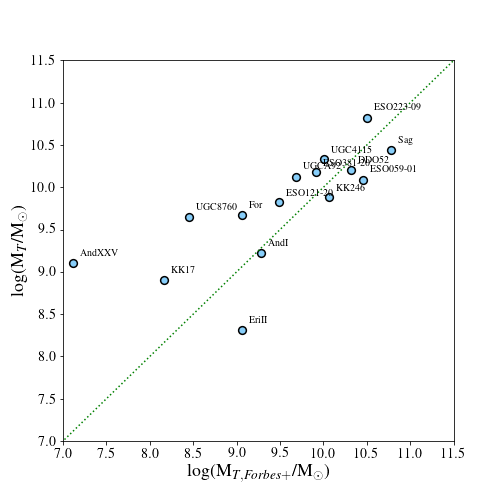}
\caption{Comparison of values of \mt\ derived using our formalism and those presented by \protect\cite{forbes18}. The most striking outlier, AndXXV, has been identified as tidally distorted \protect\citep{collins14}. The line is the 1:1 relation.}
\label{fig:mass_comp}
\end{figure}

The results of this procedure are compared to the total masses provided by \cite{forbes18} in Figure \ref{fig:mass_comp}. Once AndXXV is removed from the comparison because it is tidally distorted \citep{collins14}, the scatter is large, but workable (see \S\ref{sec:results}), particularly because it is symmetric, with an rms dispersion of 0.5 dex. 

\section{Results}
\label{sec:results}

\begin{figure}
\centering
\includegraphics[scale=0.5]{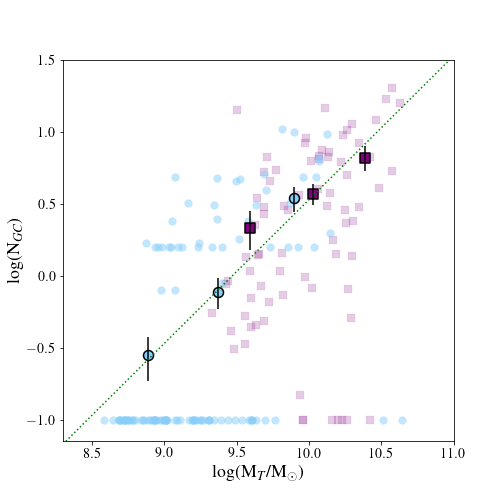}
\caption{\ngc\ vs. \mt\ for 
for the \protect\cite{forbes20} sample (squares) and \protect\cite{carlsten} sample (circles). Lightly coloured symbols without error bars are the results for individual galaxies in either sample, while the darker symbols with error bars 
bars represent means and the dispersion in the means. Individual systems with N$_{\rm GC} \le 0$ are plotted at 
$\log($\ngc$) = -1$, but the actual values are used in the calculation of the means and their uncertainties. The dotted
line represents the best-fit power law relation to the binned data.}
\label{fig:relation}
\end{figure}

From the two samples, we have \ngc\ and \mt\ for a total of 157 galaxies with \mt $\ \le 10^{10}$ M$_\odot$, a significant increase over what was previously available. 
We present the resulting \ngc-\mt\ relation in Figure \ref{fig:relation}. The Figure includes both the data for the individual galaxies in the two samples and also averages in bins of roughly 0.5 dex width in \mt. The width of these bins is broadly set by the uncertainty in the individual galaxy mass estimates (see \S\ref{sec:fm}). The average in each bin includes all \ngc\ measurements, including those with tabulated, unphysical values of N$_{\rm GC}$ that are $< 0$. The inclusion of the negative N$_{\rm GC}$ values is critical because it maintains the proper statistical behaviour of the sample. The errors on small values of N$_{\rm GC}$ are not Poissonian but rather Gaussian because the measurement is dominated by the uncertainties in the background level. The Gaussian nature of the uncertainties in the individual measurements is a necessary condition for the unbiased evaluation of the means and our use of the scatter about the mean to determine the uncertainty in the mean.

There are various results of note in the Figure. First, the averages for the \cite{forbes20} and \cite{carlsten} samples agree well. This result supports our recalibration of the \cite{forbes20} \ngc\ values, although we still caution against a detailed comparison between the two samples. Specifically, the agreement could be somewhat fortuitous and should not (yet) be used as evidence against environmental differences in the \ngc-\mt\ relation \citep{carlsten}.
Second, both sets of data together make a compelling case for the near linearity of the \ngc-\mt\ relation to as small a mass as available in the samples, \mt $\ \sim 10^{8}$ M$_\odot$. 
When we allow for a non-linear power law, N$_{\rm GC} = A$M$_{\rm T}^\beta$, our best-fit to the binned data, $\beta = 0.92\pm0.08$, is statistically consistent with a linear relation. We conclude that we find no evidence for a deviation from a linear relation down to masses of $\sim 10^{8.75}$ M$_\odot$. Third, at these limiting low masses, each galaxy has on average $\sim 0.3$ GC's, which illustrates how difficult it will be to extend the relationship to even lower masses. Fourth,
our best fit linear relation for all of the binned data is for one GC per $(2.9 \pm 0.3)\times10^9$ M$_\odot$. This result calls for a larger number of GCs per unit mass than determined previously \citep[e.g., one GC per $5\times10^9$ M$_\odot$;][]{burkert}, but the offset is sufficiently modest ($<$ a factor of two) that it could be due to either systematics in the GC completeness corrections or in the definition of \mt. The dominant uncertainties in our fitted normalisation are not included in the quoted formal uncertainty. As one example, the choice we made to adjust the value of $A$ in Eq. 2 affects the overall normalisation of the relationship and the agreement between the \cite{forbes20} and \cite{carlsten}, although the latter is also at the mercy of our recalibration of the \cite{forbes20} \ngc\ values. Fortunately, those uncertainties do not qualitatively affect any of our other key findings.

The precision of the fitted power-law exponent poses a number of challenges to models of globular cluster formation and evolution:

\medskip
\noindent
First, it unequivocally confirms the findings of \cite{forbes18} that the relation between the nature of the GC population and \mt\ extends at least 2.5 orders of magnitude in mass below where the \cite{el-badry} fiducial model relationship turns over and about 1.5 orders of magnitude in mass below the adopted limiting mass for halos that host GC formation. As noted by \cite{el-badry}, the fiducial model has various parameters that can be altered and at least the limiting halo mass could be decreased somewhat relative to the fiducial model. However, the \ngc-\mt\ relation now clearly extends well below the mass range where mergers dominate galaxy growth \citep{fitts,martin}, which argues
against the hypothesis that the relationship is simply an emergent property of hierarchical galaxy formation.

\medskip
\noindent
Second, it establishes to high precision that this relation is indeed nearly linear, despite what is expected to be a complicated interplay of GC formation and destruction phenomena present in most models \citep[e.g.,][]{ashman,elmegreen,kravtsov,kruijssen15,bastian20}. For massive galaxies, the model 'details' are thought to be mostly obscured through the homogenising effects of hierarchical growth \citep{mbk}. Indeed, \cite{el-badry} demonstrated that even a model with a random allocation of GCs to halos at early times would produce the observed relationship for galaxies built up over time from other galaxies. However, the lower mass systems included here may typically experience only a single merger \citep{martin}, if even that, and modifications in GC formation model 'details' should have results that are far more evident in the galaxies explored here.

\medskip
\noindent
Third, it emphasises the importance of the relation in terms of numbers of GCs. Most studies have focused on the \mgc-\mt\ relation, partly because \mgc\ is dominated by the more massive GCs, which are less susceptible to tidal destruction and evaporation. However, the near linearity of the \ngc-\mt\ relation reaffirms it as another useful mass proxy even for low mass galaxies. If we assume that the \mgc-\mt\ relation is the one that is perfectly linear, then we infer that the mean GC mass $\propto$ \mt$^{0.08}$, which in turn implies a $\sim$ 30\% decline in the mean GC mass going from galaxies with log(\mt) of 10.5 to 9. This is at least qualitatively consistent with the trend measured by \cite{jordan} for a different sample of galaxies. As such, we cannot yet conclude whether \ngc\ or \mgc\ is more directly tied to \mt. 

\medskip
\noindent
Fourth, it confirms GC formation as a consistent, common phenomenon across many orders in galaxy mass, even within galaxies that on average host fewer than one GC. This finding is somewhat surprising given that we expect low mass galaxies will have difficulties cooling gas and generating the extreme conditions envisioned in GC formation and evolution. Indeed, this is along the lines of the argument put forward by \cite{el-badry} for a threshold halo mass in their model. GC formation models tend to posit the creation of high pressure, high density gas disks at large redshifts \citep{elmegreen,kravtsov,kruijssen15} in which GCs form and may be tidally destroyed before mergers scatter the GCs into the galaxy halo \citep{kruijssen15}. This process would have to have a nearly constant efficiency across at least 4 orders of halo mass to reproduce the observed relationships (7 orders of magnitude if these results can be grafted onto those discussed by \cite{burkert}). Furthermore, any scenario that involves redistribution of GCs by mergers faces a related challenge in that the GC distributions in these low mass systems are not disk-like and closely resemble the underlying stellar distributions in both their radial and azimuthal distributions \citep{carlsten,teymoor2}.

From a different perspective, 
the near linearity of the \ngc-\mt\ relation supports the use of the galaxy scaling relations for determining \mt. While \ngc\ has been used as a mass proxy \citep[e.g,][]{vdk16,amorisco18,forbes20}, measuring \ngc\ is also observationally challenging. The use of the scaling relation methodology will open up far larger samples for study.

A contrasting opinion is presented by \cite{bastian20} who emphasise that the masses assigned to these low luminosity galaxies in GC studies are in conflict with those derived from the stellar mass-halo mass relation, M$_*$-\mt. They favour using the halo masses estimated form M$_*$-\mt, which do reproduce the downturn in the \mgc-\mt\ relation produced by their simulations. Thus, there is a clear tension between dynamically-estimated masses \cite[such as those estimated here and by][]{forbes20} for low mass galaxies and those estimated using M$_*$-\mt. 
The fault is not attributed to the determination of the mass within $r_e$ but rather in the extrapolation to \mt. The discrepancies can be larger than an order of magnitude \citep{bastian20}, which seem to us difficult to reconcile with our rather direct use of NFW profiles to estimate \mt. 
In further support of our determination of \mt\ we note that a systematic error in \mt\ estimation would be unlikely to result in the observed precisely linear \ngc-\mt\ relation. For example, replacing our estimates of M$_{\rm T}$ with those derived from the \cite{moster} M$_{*}$-M$_{\rm T}$ relation using the parameters they derive for all galaxies at $z=0.1$ results in $\beta = 1.83 \pm 0.36$, which is a 2.3$\sigma$ deviation from linearity. As such, the tension remains and is an interesting avenue for further investigation.

\section{Summary}

Using a new method with which to estimate the total mass of galaxies and two existing large sample of galaxies with measurements of the number of globular clusters \citep{forbes20,carlsten}, we revisit the relationship between the number of globular clusters (GCs) a galaxy hosts, \ngc, and its total mass, \mt. We confirm previous findings \citep{forbes18} that the relation extends to low \mt, $\sim 10^{8.75}$ M$_\odot$, and use these larger samples to place tight constraints on the power law nature of the relationship, \ngc\ $\propto $ \mt$^{0.92\pm0.08}$. For an adopted linear relation, we find one GC per  $(2.9 \pm 0.3)\times10^9$ M$_\odot$ of total galaxy mass, although this number has significant additional uncertainty beyond the formal quoted uncertainty that is related to the adopted GC completeness corrections and estimation of \mt.

The nearly linear behaviour of the relationship places strong constraints on models, particularly since this behaviour extends to galaxy masses well below where galaxies have less than one GC on average. Any potential threshold halo mass for the production of GCs lies below 10$^{8.75}$ M$_\odot$, and the correlation between \ngc\ and \mt, at least at the lowest masses, cannot be an emergent property of the hierarchical growth of galaxies. Models that invoke mergers and interactions as a means either to trigger cluster formation or to redistribute clusters into the galaxy halos \citep[e.g.,][]{kruijssen15} will also face challenges because the \ngc-\mt\ relation extends to masses where mergers are uncommon \citep{fitts,martin}. 

We believe that these results motivate the investigation of novel GC formation models, such as the one proposed based on streaming motions of baryons in the early universe \citep{naoz,schauer}. Extending the \ngc-\mt\ relation to even lower masses is a challenge, but there are indications that even ultrafaint galaxies can host GCs \citep{crnojevic} and that on average these galaxies continue the \ngc-\mt\ relation to even lower masses \citep{zar16}. 
GCs have played an outsized role in our understanding of the Universe, ranging from helping us to uncover the nature of our own galaxy \citep{shapley} and others \citep{hubble} to developing a theory of the evolution of stars and stellar populations \citep{sandage}. They may yet reveal additional fundamental astrophysics.

\section*{acknowledgments}
DZ acknowledges financial support from and AST-2006785, Duncan Forbes, Scott Carlsten, and the anonymous referee for helpful comments, and Scott Carlsten for sharing his sample prior to publication.

%\software{
%Astropy              \citep{astropy1, astropy2},
%astroquery           \citep{astroquery},
%GALFIT               \citep{peng},
%keras                \citep{keras},
%lmfit                \citep{newville},
%Matplotlib           \citep{matplotlib},
%NumPy                \citep{numpy},
%pandas               \citep{pandas},
%sep                  \citep{sep},
%Source Extractor     \citep{bertin},
%SciPy                \citep{scipy1, scipy2},
%SWarp                \citep{Swarp}
%}

\section*{Data Availability}
No new data were generated or analysed in support of this research.



\bibliographystyle{mnras}
\bibliography{zaritsky.bib} % if your bibtex file is called example.bib

\begin{thebibliography}{}
\makeatletter
\relax
\def\mn@urlcharsother{\let\do\@makeother \do\$\do\&\do\#\do\^\do\_\do\%\do\~}
\def\mn@doi{\begingroup\mn@urlcharsother \@ifnextchar [ {\mn@doi@}
  {\mn@doi@[]}}
\def\mn@doi@[#1]#2{\def\@tempa{#1}\ifx\@tempa\@empty \href
  {http://dx.doi.org/#2} {doi:#2}\else \href {http://dx.doi.org/#2} {#1}\fi
  \endgroup}
\def\mn@eprint#1#2{\mn@eprint@#1:#2::\@nil}
\def\mn@eprint@arXiv#1{\href {http://arxiv.org/abs/#1} {{\tt arXiv:#1}}}
\def\mn@eprint@dblp#1{\href {http://dblp.uni-trier.de/rec/bibtex/#1.xml}
  {dblp:#1}}
\def\mn@eprint@#1:#2:#3:#4\@nil{\def\@tempa {#1}\def\@tempb {#2}\def\@tempc
  {#3}\ifx \@tempc \@empty \let \@tempc \@tempb \let \@tempb \@tempa \fi \ifx
  \@tempb \@empty \def\@tempb {arXiv}\fi \@ifundefined
  {mn@eprint@\@tempb}{\@tempb:\@tempc}{\expandafter \expandafter \csname
  mn@eprint@\@tempb\endcsname \expandafter{\@tempc}}}

\bibitem[\protect\citeauthoryear{{Alabi}, {Romanowsky}, {Forbes}, {Brodie}  \&
  {Okabe}}{{Alabi} et~al.}{2020}]{alabi}
{Alabi} A.~B.,  {Romanowsky} A.~J.,  {Forbes} D.~A.,  {Brodie} J.~P.,   {Okabe}
  N.,  2020, \mn@doi [\mnras] {10.1093/mnras/staa1763}, \href
  {https://ui.adsabs.harvard.edu/abs/2020MNRAS.496.3182A} {496, 3182}

\bibitem[\protect\citeauthoryear{{Amorisco}, {Monachesi}, {Agnello}  \&
  {White}}{{Amorisco} et~al.}{2018}]{amorisco18}
{Amorisco} N.~C.,  {Monachesi} A.,  {Agnello} A.,   {White} S.~D.~M.,  2018,
  \mn@doi [\mnras] {10.1093/mnras/sty116}, \href
  {https://ui.adsabs.harvard.edu/abs/2018MNRAS.475.4235A} {475, 4235}

\bibitem[\protect\citeauthoryear{{Ashman} \& {Zepf}}{{Ashman} \&
  {Zepf}}{1992}]{ashman}
{Ashman} K.~M.,  {Zepf} S.~E.,  1992, \mn@doi [\apj] {10.1086/170850}, \href
  {https://ui.adsabs.harvard.edu/abs/1992ApJ...384...50A} {384, 50}

\bibitem[\protect\citeauthoryear{{Bastian}, {Pfeffer}, {Kruijssen}, {Crain},
  {Trujillo-Gomez}  \& {Reina-Campos}}{{Bastian} et~al.}{2020}]{bastian20}
{Bastian} N.,  {Pfeffer} J.,  {Kruijssen} J.~M.~D.,  {Crain} R.~A.,
  {Trujillo-Gomez} S.,   {Reina-Campos} M.,  2020, \mn@doi [\mnras]
  {10.1093/mnras/staa2453}, \href
  {https://ui.adsabs.harvard.edu/abs/2020MNRAS.498.1050B} {498, 1050}

\bibitem[\protect\citeauthoryear{{Blakeslee}, {Tonry}  \&
  {Metzger}}{{Blakeslee} et~al.}{1997}]{blakeslee}
{Blakeslee} J.~P.,  {Tonry} J.~L.,   {Metzger} M.~R.,  1997, \mn@doi [\aj]
  {10.1086/118488}, \href {http://adsabs.harvard.edu/abs/1997AJ....114..482B}
  {114, 482}

\bibitem[\protect\citeauthoryear{{Bovy}}{{Bovy}}{2015}]{bovy}
{Bovy} J.,  2015, \mn@doi [\apjs] {10.1088/0067-0049/216/2/29}, \href
  {https://ui.adsabs.harvard.edu/abs/2015ApJS..216...29B} {216, 29}

\bibitem[\protect\citeauthoryear{{Boylan-Kolchin}}{{Boylan-Kolchin}}{2017}]{mbk}
{Boylan-Kolchin} M.,  2017, \mn@doi [\mnras] {10.1093/mnras/stx2164}, \href
  {https://ui.adsabs.harvard.edu/abs/2017MNRAS.472.3120B} {472, 3120}

\bibitem[\protect\citeauthoryear{{Bryan} \& {Norman}}{{Bryan} \&
  {Norman}}{1998}]{bryan}
{Bryan} G.~L.,  {Norman} M.~L.,  1998, \mn@doi [\apj] {10.1086/305262}, \href
  {https://ui.adsabs.harvard.edu/abs/1998ApJ...495...80B} {495, 80}

\bibitem[\protect\citeauthoryear{{Burkert} \& {Forbes}}{{Burkert} \&
  {Forbes}}{2020}]{burkert}
{Burkert} A.,  {Forbes} D.~A.,  2020, \mn@doi [\aj] {10.3847/1538-3881/ab5b0e},
  \href {https://ui.adsabs.harvard.edu/abs/2020AJ....159...56B} {159, 56}

\bibitem[\protect\citeauthoryear{{Carlsten}, {Greene}, {Beaton}  \&
  {Greco}}{{Carlsten} et~al.}{2021a}]{carlsten}
{Carlsten} S.~G.,  {Greene} J.~E.,  {Beaton} R.~L.,   {Greco} J.~P.,  2021a,
  arXiv e-prints, \href {https://ui.adsabs.harvard.edu/abs/2021arXiv210503440C}
  {p. arXiv:2105.03440}

\bibitem[\protect\citeauthoryear{{Carlsten}, {Greene}, {Greco}, {Beaton}  \&
  {Kado-Fong}}{{Carlsten} et~al.}{2021b}]{carlsten21}
{Carlsten} S.~G.,  {Greene} J.~E.,  {Greco} J.~P.,  {Beaton} R.~L.,
  {Kado-Fong} E.,  2021b, \mn@doi [\apj] {10.3847/1538-4357/ac2581}, \href
  {https://ui.adsabs.harvard.edu/abs/2021ApJ...922..267C} {922, 267}

\bibitem[\protect\citeauthoryear{{Choksi} \& {Gnedin}}{{Choksi} \&
  {Gnedin}}{2019}]{choksi}
{Choksi} N.,  {Gnedin} O.~Y.,  2019, \mn@doi [\mnras] {10.1093/mnras/stz2097},
  \href {https://ui.adsabs.harvard.edu/abs/2019MNRAS.488.5409C} {488, 5409}

\bibitem[\protect\citeauthoryear{{Collins} et~al.,}{{Collins}
  et~al.}{2013}]{collins13}
{Collins} M. L.~M.,  et~al., 2013, \mn@doi [\apj]
  {10.1088/0004-637X/768/2/172}, \href
  {https://ui.adsabs.harvard.edu/abs/2013ApJ...768..172C} {768, 172}

\bibitem[\protect\citeauthoryear{{Collins} et~al.,}{{Collins}
  et~al.}{2014}]{collins14}
{Collins} M. L.~M.,  et~al., 2014, \mn@doi [\apj] {10.1088/0004-637X/783/1/7},
  \href {https://ui.adsabs.harvard.edu/abs/2014ApJ...783....7C} {783, 7}

\bibitem[\protect\citeauthoryear{{Crnojevi{\'c}}, {Sand}, {Zaritsky},
  {Spekkens}, {Willman}  \& {Hargis}}{{Crnojevi{\'c}} et~al.}{2016}]{crnojevic}
{Crnojevi{\'c}} D.,  {Sand} D.~J.,  {Zaritsky} D.,  {Spekkens} K.,  {Willman}
  B.,   {Hargis} J.~R.,  2016, \mn@doi [\apjl] {10.3847/2041-8205/824/1/L14},
  \href {https://ui.adsabs.harvard.edu/abs/2016ApJ...824L..14C} {824, L14}

\bibitem[\protect\citeauthoryear{{Djorgovski} \& {Davis}}{{Djorgovski} \&
  {Davis}}{1987}]{djorgovski}
{Djorgovski} S.,  {Davis} M.,  1987, \mn@doi [\apj] {10.1086/164948}, \href
  {https://ui.adsabs.harvard.edu/abs/1987ApJ...313...59D} {313, 59}

\bibitem[\protect\citeauthoryear{{\VAN{Dokkum}{van}{van} Dokkum}
  et~al.,}{{\VAN{Dokkum}{van}{van} Dokkum} et~al.}{2016}]{vdk16}
{\VAN{Dokkum}{van}{van} Dokkum} P.,  et~al., 2016, \mn@doi [\apjl]
  {10.3847/2041-8205/828/1/L6}, \href
  {https://ui.adsabs.harvard.edu/abs/2016ApJ...828L...6V} {828, L6}

\bibitem[\protect\citeauthoryear{{\VAN{Dokkum}{van}{van} Dokkum}
  et~al.,}{{\VAN{Dokkum}{van}{van} Dokkum} et~al.}{2017}]{vdk17}
{\VAN{Dokkum}{van}{van} Dokkum} P.,  et~al., 2017, \mn@doi [\apjl]
  {10.3847/2041-8213/aa7ca2}, \href
  {https://ui.adsabs.harvard.edu/abs/2017ApJ...844L..11V} {844, L11}

\bibitem[\protect\citeauthoryear{{Dressler}, {Lynden-Bell}, {Burstein},
  {Davies}, {Faber}, {Terlevich}  \& {Wegner}}{{Dressler}
  et~al.}{1987}]{dressler}
{Dressler} A.,  {Lynden-Bell} D.,  {Burstein} D.,  {Davies} R.~L.,  {Faber}
  S.~M.,  {Terlevich} R.,   {Wegner} G.,  1987, \mn@doi [\apj]
  {10.1086/164947}, \href
  {https://ui.adsabs.harvard.edu/abs/1987ApJ...313...42D} {313, 42}

\bibitem[\protect\citeauthoryear{{Eadie}, {Harris}  \& {Springford}}{{Eadie}
  et~al.}{2022}]{eadie}
{Eadie} G.~M.,  {Harris} W.~E.,   {Springford} A.,  2022, \mn@doi [\apj]
  {10.3847/1538-4357/ac33b0}, \href
  {https://ui.adsabs.harvard.edu/abs/2022ApJ...926..162E} {926, 162}

\bibitem[\protect\citeauthoryear{{El-Badry}, {Quataert}, {Weisz}, {Choksi}  \&
  {Boylan-Kolchin}}{{El-Badry} et~al.}{2019}]{el-badry}
{El-Badry} K.,  {Quataert} E.,  {Weisz} D.~R.,  {Choksi} N.,   {Boylan-Kolchin}
  M.,  2019, \mn@doi [\mnras] {10.1093/mnras/sty3007}, \href
  {http://adsabs.harvard.edu/abs/2019MNRAS.482.4528E} {482, 4528}

\bibitem[\protect\citeauthoryear{{Elmegreen} \& {Efremov}}{{Elmegreen} \&
  {Efremov}}{1997}]{elmegreen}
{Elmegreen} B.~G.,  {Efremov} Y.~N.,  1997, \mn@doi [\apj] {10.1086/303966},
  \href {https://ui.adsabs.harvard.edu/abs/1997ApJ...480..235E} {480, 235}

\bibitem[\protect\citeauthoryear{{Fitts} et~al.,}{{Fitts} et~al.}{2018}]{fitts}
{Fitts} A.,  et~al., 2018, \mn@doi [\mnras] {10.1093/mnras/sty1488}, \href
  {https://ui.adsabs.harvard.edu/abs/2018MNRAS.479..319F} {479, 319}

\bibitem[\protect\citeauthoryear{{Forbes}, {Alabi}, {Romanowsky}, {Brodie},
  {Strader}, {Usher}  \& {Pota}}{{Forbes} et~al.}{2016}]{forbes16}
{Forbes} D.~A.,  {Alabi} A.,  {Romanowsky} A.~J.,  {Brodie} J.~P.,  {Strader}
  J.,  {Usher} C.,   {Pota} V.,  2016, \mn@doi [\mnras]
  {10.1093/mnrasl/slw015}, \href
  {http://adsabs.harvard.edu/abs/2016MNRAS.458L..44F} {458, L44}

\bibitem[\protect\citeauthoryear{{Forbes}, {Read}, {Gieles}  \&
  {Collins}}{{Forbes} et~al.}{2018}]{forbes18}
{Forbes} D.~A.,  {Read} J.~I.,  {Gieles} M.,   {Collins} M. L.~M.,  2018,
  \mn@doi [\mnras] {10.1093/mnras/sty2584}, \href
  {https://ui.adsabs.harvard.edu/abs/2018MNRAS.481.5592F} {481, 5592}

\bibitem[\protect\citeauthoryear{{Forbes}, {Alabi}, {Romanowsky}, {Brodie}  \&
  {Arimoto}}{{Forbes} et~al.}{2020}]{forbes20}
{Forbes} D.~A.,  {Alabi} A.,  {Romanowsky} A.~J.,  {Brodie} J.~P.,   {Arimoto}
  N.,  2020, \mn@doi [\mnras] {10.1093/mnras/staa180}, \href
  {https://ui.adsabs.harvard.edu/abs/2020MNRAS.492.4874F} {492, 4874}

\bibitem[\protect\citeauthoryear{{Georgiev}, {Puzia}, {Goudfrooij}  \&
  {Hilker}}{{Georgiev} et~al.}{2010}]{georgiev}
{Georgiev} I.~Y.,  {Puzia} T.~H.,  {Goudfrooij} P.,   {Hilker} M.,  2010,
  \mn@doi [\mnras] {10.1111/j.1365-2966.2010.16802.x}, \href
  {http://adsabs.harvard.edu/abs/2010MNRAS.406.1967G} {406, 1967}

\bibitem[\protect\citeauthoryear{{Harris}, {Harris}  \& {Alessi}}{{Harris}
  et~al.}{2013}]{harris13}
{Harris} W.~E.,  {Harris} G.~L.~H.,   {Alessi} M.,  2013, \mn@doi [\apj]
  {10.1088/0004-637X/772/2/82}, \href
  {http://adsabs.harvard.edu/abs/2013ApJ...772...82H} {772, 82}

\bibitem[\protect\citeauthoryear{{Harris}, {Blakeslee}  \& {Harris}}{{Harris}
  et~al.}{2017}]{harris17}
{Harris} W.~E.,  {Blakeslee} J.~P.,   {Harris} G.~L.~H.,  2017, \mn@doi [\apj]
  {10.3847/1538-4357/836/1/67}, \href
  {http://adsabs.harvard.edu/abs/2017ApJ...836...67H} {836, 67}

\bibitem[\protect\citeauthoryear{{Hinshaw} et~al.,}{{Hinshaw}
  et~al.}{2013}]{wmap}
{Hinshaw} G.,  et~al., 2013, \mn@doi [\apjs] {10.1088/0067-0049/208/2/19},
  \href {https://ui.adsabs.harvard.edu/abs/2013ApJS..208...19H} {208, 19}

\bibitem[\protect\citeauthoryear{{Hubble}}{{Hubble}}{1932}]{hubble}
{Hubble} E.,  1932, \mn@doi [\apj] {10.1086/143397}, \href
  {http://adsabs.harvard.edu/abs/1932ApJ....76...44H} {76, 44}

\bibitem[\protect\citeauthoryear{{Hudson}, {Harris}  \& {Harris}}{{Hudson}
  et~al.}{2014}]{hudson}
{Hudson} M.~J.,  {Harris} G.~L.,   {Harris} W.~E.,  2014, \mn@doi [\apjl]
  {10.1088/2041-8205/787/1/L5}, \href
  {https://ui.adsabs.harvard.edu/abs/2014ApJ...787L...5H} {787, L5}

\bibitem[\protect\citeauthoryear{{Jord{\'a}n} et~al.,}{{Jord{\'a}n}
  et~al.}{2006}]{jordan}
{Jord{\'a}n} A.,  et~al., 2006, \mn@doi [\apjl] {10.1086/509119}, \href
  {https://ui.adsabs.harvard.edu/abs/2006ApJ...651L..25J} {651, L25}

\bibitem[\protect\citeauthoryear{{Kravtsov} \& {Gnedin}}{{Kravtsov} \&
  {Gnedin}}{2005}]{kravtsov}
{Kravtsov} A.~V.,  {Gnedin} O.~Y.,  2005, \mn@doi [\apj] {10.1086/428636},
  \href {https://ui.adsabs.harvard.edu/abs/2005ApJ...623..650K} {623, 650}

\bibitem[\protect\citeauthoryear{{Kruijssen}}{{Kruijssen}}{2015}]{kruijssen15}
{Kruijssen} J.~M.~D.,  2015, \mn@doi [\mnras] {10.1093/mnras/stv2026}, \href
  {https://ui.adsabs.harvard.edu/abs/2015MNRAS.454.1658K} {454, 1658}

\bibitem[\protect\citeauthoryear{{Lim}, {Peng}, {C{\^o}t{\'e}}, {Sales}, {den
  Brok}, {Blakeslee}  \& {Guhathakurta}}{{Lim} et~al.}{2018}]{lim}
{Lim} S.,  {Peng} E.~W.,  {C{\^o}t{\'e}} P.,  {Sales} L.~V.,  {den Brok} M.,
  {Blakeslee} J.~P.,   {Guhathakurta} P.,  2018, \mn@doi [\apj]
  {10.3847/1538-4357/aacb81}, \href
  {https://ui.adsabs.harvard.edu/abs/2018ApJ...862...82L} {862, 82}

\bibitem[\protect\citeauthoryear{{Macci{\`o}}, {Dutton}, {van den Bosch},
  {Moore}, {Potter}  \& {Stadel}}{{Macci{\`o}} et~al.}{2007}]{maccio}
{Macci{\`o}} A.~V.,  {Dutton} A.~A.,  {van den Bosch} F.~C.,  {Moore} B.,
  {Potter} D.,   {Stadel} J.,  2007, \mn@doi [\mnras]
  {10.1111/j.1365-2966.2007.11720.x}, \href
  {https://ui.adsabs.harvard.edu/abs/2007MNRAS.378...55M} {378, 55}

\bibitem[\protect\citeauthoryear{{Martin} et~al.,}{{Martin}
  et~al.}{2016}]{martin16}
{Martin} N.~F.,  et~al., 2016, \mn@doi [\apj] {10.3847/1538-4357/833/2/167},
  \href {https://ui.adsabs.harvard.edu/abs/2016ApJ...833..167M} {833, 167}

\bibitem[\protect\citeauthoryear{{Martin} et~al.,}{{Martin}
  et~al.}{2021}]{martin}
{Martin} G.,  et~al., 2021, \mn@doi [\mnras] {10.1093/mnras/staa3443}, \href
  {https://ui.adsabs.harvard.edu/abs/2021MNRAS.500.4937M} {500, 4937}

\bibitem[\protect\citeauthoryear{{McConnachie}}{{McConnachie}}{2012}]{mcc}
{McConnachie} A.~W.,  2012, \mn@doi [\aj] {10.1088/0004-6256/144/1/4}, \href
  {https://ui.adsabs.harvard.edu/abs/2012AJ....144....4M} {144, 4}

\bibitem[\protect\citeauthoryear{{McGaugh} \& {Schombert}}{{McGaugh} \&
  {Schombert}}{2014}]{mcgaugh}
{McGaugh} S.~S.,  {Schombert} J.~M.,  2014, \mn@doi [\aj]
  {10.1088/0004-6256/148/5/77}, \href
  {https://ui.adsabs.harvard.edu/abs/2014AJ....148...77M} {148, 77}

\bibitem[\protect\citeauthoryear{{Moster}, {Naab}  \& {White}}{{Moster}
  et~al.}{2018}]{moster}
{Moster} B.~P.,  {Naab} T.,   {White} S. D.~M.,  2018, \mn@doi [\mnras]
  {10.1093/mnras/sty655}, \href
  {https://ui.adsabs.harvard.edu/abs/2018MNRAS.477.1822M} {477, 1822}

\bibitem[\protect\citeauthoryear{{Naoz} \& {Narayan}}{{Naoz} \&
  {Narayan}}{2014}]{naoz}
{Naoz} S.,  {Narayan} R.,  2014, \mn@doi [\apjl] {10.1088/2041-8205/791/1/L8},
  \href {https://ui.adsabs.harvard.edu/abs/2014ApJ...791L...8N} {791, L8}

\bibitem[\protect\citeauthoryear{{Navarro}, {Frenk}  \& {White}}{{Navarro}
  et~al.}{1997}]{navarro}
{Navarro} J.~F.,  {Frenk} C.~S.,   {White} S.~D.~M.,  1997, \mn@doi [\apj]
  {10.1086/304888}, \href {http://adsabs.harvard.edu/abs/1997ApJ...490..493N}
  {490, 493}

\bibitem[\protect\citeauthoryear{{Saifollahi} et~al.,}{{Saifollahi}
  et~al.}{2022}]{teymoor2}
{Saifollahi} T.,  et~al., 2022, \mnras

\bibitem[\protect\citeauthoryear{{Sandage} \& {Schwarzschild}}{{Sandage} \&
  {Schwarzschild}}{1952}]{sandage}
{Sandage} A.~R.,  {Schwarzschild} M.,  1952, \mn@doi [\apj] {10.1086/145638},
  \href {http://adsabs.harvard.edu/abs/1952ApJ...116..463S} {116, 463}

\bibitem[\protect\citeauthoryear{{Schauer}, {Bromm}, {Boylan-Kolchin}, {Glover}
   \& {Klessen}}{{Schauer} et~al.}{2021}]{schauer}
{Schauer} A. T.~P.,  {Bromm} V.,  {Boylan-Kolchin} M.,  {Glover} S. C.~O.,
  {Klessen} R.~S.,  2021, \mn@doi [\apj] {10.3847/1538-4357/ac27aa}, \href
  {https://ui.adsabs.harvard.edu/abs/2021ApJ...922..193S} {922, 193}

\bibitem[\protect\citeauthoryear{{Shapley}}{{Shapley}}{1918}]{shapley}
{Shapley} H.,  1918, \mn@doi [\apj] {10.1086/142423}, \href
  {http://adsabs.harvard.edu/abs/1918ApJ....48..154S} {48}

\bibitem[\protect\citeauthoryear{{Spitler} \& {Forbes}}{{Spitler} \&
  {Forbes}}{2009}]{spitler}
{Spitler} L.~R.,  {Forbes} D.~A.,  2009, \mn@doi [\mnras]
  {10.1111/j.1745-3933.2008.00567.x}, \href
  {http://adsabs.harvard.edu/abs/2009MNRAS.392L...1S} {392, L1}

\bibitem[\protect\citeauthoryear{{Tollerud} et~al.,}{{Tollerud}
  et~al.}{2012}]{tollerud12}
{Tollerud} E.~J.,  et~al., 2012, \mn@doi [\apj] {10.1088/0004-637X/752/1/45},
  \href {https://ui.adsabs.harvard.edu/abs/2012ApJ...752...45T} {752, 45}

\bibitem[\protect\citeauthoryear{{Tollerud}, {Geha}, {Vargas}  \&
  {Bullock}}{{Tollerud} et~al.}{2013}]{tollerud13}
{Tollerud} E.~J.,  {Geha} M.~C.,  {Vargas} L.~C.,   {Bullock} J.~S.,  2013,
  \mn@doi [\apj] {10.1088/0004-637X/768/1/50}, \href
  {https://ui.adsabs.harvard.edu/abs/2013ApJ...768...50T} {768, 50}

\bibitem[\protect\citeauthoryear{{Willmer}}{{Willmer}}{2018}]{willmer}
{Willmer} C. N.~A.,  2018, \mn@doi [\apjs] {10.3847/1538-4365/aabfdf}, \href
  {https://ui.adsabs.harvard.edu/abs/2018ApJS..236...47W} {236, 47}

\bibitem[\protect\citeauthoryear{{Zaritsky}}{{Zaritsky}}{2012}]{zaritsky12}
{Zaritsky} D.,  2012, \mn@doi [ISRN Astronomy and Astrophysics]
  {10.5402/2012/189625}, \href
  {https://ui.adsabs.harvard.edu/abs/2012ISRAA2012E..12Z} {2012, 189625}

\bibitem[\protect\citeauthoryear{{Zaritsky}}{{Zaritsky}}{2017}]{zar17}
{Zaritsky} D.,  2017, \mn@doi [\mnras] {10.1093/mnrasl/slw198}, \href
  {https://ui.adsabs.harvard.edu/abs/2017MNRAS.464L.110Z} {464, L110}

\bibitem[\protect\citeauthoryear{{Zaritsky}, {Gonzalez}  \&
  {Zabludoff}}{{Zaritsky} et~al.}{2006a}]{FM}
{Zaritsky} D.,  {Gonzalez} A.~H.,   {Zabludoff} A.~I.,  2006a, \mn@doi [\apj]
  {10.1086/498672}, \href
  {https://ui.adsabs.harvard.edu/abs/2006ApJ...638..725Z} {638, 725}

\bibitem[\protect\citeauthoryear{{Zaritsky}, {Gonzalez}  \&
  {Zabludoff}}{{Zaritsky} et~al.}{2006b}]{FM-LG}
{Zaritsky} D.,  {Gonzalez} A.~H.,   {Zabludoff} A.~I.,  2006b, \mn@doi [\apjl]
  {10.1086/504352}, \href
  {https://ui.adsabs.harvard.edu/abs/2006ApJ...642L..37Z} {642, L37}

\bibitem[\protect\citeauthoryear{{Zaritsky}, {Zabludoff}  \&
  {Gonzalez}}{{Zaritsky} et~al.}{2008}]{FM-structure}
{Zaritsky} D.,  {Zabludoff} A.~I.,   {Gonzalez} A.~H.,  2008, \mn@doi [\apj]
  {10.1086/529577}, \href
  {https://ui.adsabs.harvard.edu/abs/2008ApJ...682...68Z} {682, 68}

\bibitem[\protect\citeauthoryear{{Zaritsky}, {Zabludoff}  \&
  {Gonzalez}}{{Zaritsky} et~al.}{2011}]{FM-clusters}
{Zaritsky} D.,  {Zabludoff} A.~I.,   {Gonzalez} A.~H.,  2011, \mn@doi [\apj]
  {10.1088/0004-637X/727/2/116}, \href
  {https://ui.adsabs.harvard.edu/abs/2011ApJ...727..116Z} {727, 116}

\bibitem[\protect\citeauthoryear{{Zaritsky}, {Crnojevi{\'c}}  \&
  {Sand}}{{Zaritsky} et~al.}{2016}]{zar16}
{Zaritsky} D.,  {Crnojevi{\'c}} D.,   {Sand} D.~J.,  2016, \mn@doi [\apjl]
  {10.3847/2041-8205/826/1/L9}, \href
  {https://ui.adsabs.harvard.edu/abs/2016ApJ...826L...9Z} {826, L9}

\makeatother
\end{thebibliography}

\bsp
\label{lastpage}
\end{document}